\documentclass[aps,pra,twocolumn,showpacs,letterpaper,tighten,float,nofootinbib]{revtex4-1}
\usepackage{graphicx}
\usepackage{latexsym}
\usepackage{amsmath,amssymb}
\usepackage{bm} 
\usepackage{color}
\usepackage{epsfig}

\DeclareMathOperator{\sgn}{sgn}

\definecolor{myblue}{rgb}{.93, .93, 1}

\newcommand{\bsub}{\begin{subequations}}
	\newcommand{\esub}{\end{subequations}}

\newcommand{\vex}[1]{\bm{\mathrm{#1}}}

\newcommand{\xv}{{\bf x}}

\begin{document}
	
	\title{Localized surfaces of three dimensional topological insulators}
	\author{Yang-Zhi~Chou}\email{YangZhi.Chou@colorado.edu}  \author{Rahul~M.~Nandkishore} \author{Leo~Radzihovsky}
	
	\affiliation{Department of Physics and Center for Theory of Quantum
		Matter, University of Colorado Boulder, Boulder, Colorado 80309,
		USA} \date{\today}
	\begin{abstract}
		We study the surface of a three-dimensional spin chiral $\mathrm{Z}_2$
		topological insulator (class CII), demonstrating the possibility of
		its localization. This arises through an interplay of interaction
		and statistically-symmetric disorder, that confines 
		the gapless fermionic degrees of freedom to a network of one-dimensional \emph{helical} domain-walls 
		that can be localized. We identify two distinct regimes of this gapless
		insulating phase, a ``clogged'' regime wherein the network
		localization is induced by its junctions between otherwise metallic
		helical domain-walls, and a ``fully localized'' regime of localized
		domain-walls. The experimental signatures of these regimes are also
		discussed.
	\end{abstract}
	
	\maketitle
	
	\section{Introduction}
	
	The surfaces of topological insulators (TIs)
	\cite{Hasan2010_RMP,Qi2011_RMP,Chiu2016,Ludwig2015} exhibit robust
	symmetry-protected metallic transport even in the presence of
	symmetry-preserving heterogeneity (disorder) as long as the bulk
	remains gapped. The evasion of Anderson localization
	\cite{pwa1958,Evers2008} is due to the anomalous nature of the surface
	states, reflecting a nontrivial wavefunction topology of TI's
	bulk. Characterization of such symmetry protected topological 
	materials is a vibrant field of research in modern condensed matter
	physics \cite{Ando2015,Mizushima2016review}.
	
	Interactions can destabilize such metallic
	surfaces \cite{Vishwanath2013,Wang2013,Wang2014,Metlitski2014,SenthilARCMP},
	gapping them out by either spontaneously breaking the protective
	symmetry, or inducing a symmetry-preserving topologically-ordered
	long-range entanglement. However, it has been
	noted \cite{Xu2006,Wu2006} and explored more extensively by
	us \cite{Chou2018}, that in a two-dimensional (2D) time-reversal
	symmetric $\mathrm{Z}_2$ TI (class AII) \cite{Kane2005_1,Kane2005_2,Bernevig2006}
	an interplay of interaction and disorder can lead to another
	possibility, namely to a glassy gapless but {\em insulating}
	edge. Such a localized state breaks the time-reversal symmetry
	spontaneously, but in ``spin glass'' fashion, preserving it
	statistically. It exhibits a localization length that is a
	non-monotonic function of disorder strength, and is best viewed as a
	localized insulator of half charge fermionic domain-walls
	(Luther-Emery \cite{Luther1974} fermons) \cite{Chou2018}.  Such edge
	localization provides a potential explanation of the puzzling
	experimental observations in InAs/GaSb TI systems
	\cite{Du2015,Li2015,Du2017,Li2017}.
	
	\begin{figure}[t!]
		\includegraphics[width=0.4\textwidth]{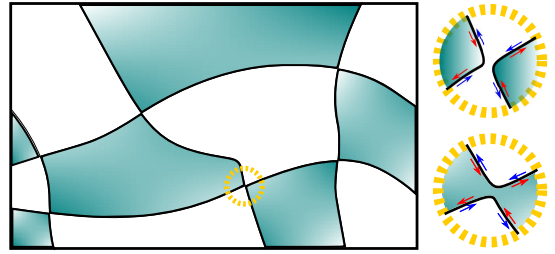}
		\caption{An illustration of a disordered interacting class CII
			TI surface, forming helical domain-walls (black solid
			curves), between topologically gapped green
			domains \cite{Morimoto2015_AL} and trivially gapped white
			regions.  Zoom-in: The interdomain four-way junction
			modeled as two helical Luttinger liquids with an impurity
			(junction) perturbation.}
		\label{Fig:Domains}
	\end{figure}
	
	Motivated by this nontrivial disorder-interaction interplay in an edge
	of a 2D TI, we explore such phenomena in a 2D surface of a three-dimensional (3D) TI 
	and find that only the CII class realizes this idea, namely, exhibits a \emph{gapless localized} surface.
	We thus focus on the CII class TIs in the presence of symmetry
	breaking, but statistically preserving disorder. Such a disorder potential can in principle be generated dynamically \cite{Morimoto2015_breakdown,Song2017}. It allows for three
	distinct possibilities: a network of chiral (particle-hole symmetric)
	or helical (time-reversal symmetric) domain-walls \cite{Potter2017}
	(see Fig.~\ref{Fig:Domains}), or a fully gapped (time-reversal and particle-hole
	symmetry-broken) insulators, depending on which symmetries are broken
	by disorder. As we demonstrate below, for the second case of a network
	of helical domain-walls, in the presence of interactions, a CII class
	TI surface indeed displays a phase transition to a {\em gapless
		insulating} surface. The latter exhibits two regimes: a ``clogged''
	regime in which the barriers to transport are the junctions in the
	network of otherwise delocalized domain-walls \cite{Teo2009}, and a
	fully localized regime of interpenetrating one-dimensional (1D) localized helical edge
	states \cite{Chou2018}. These interaction-induced regimes are obtained 
	via standard analysis for helical Luttinger liquids \cite{Wu2006,Xu2006,Chou2018,Teo2009}. Topological insulators in other symmetry
	classes of the ten-fold way do not allow this novel possibility.
	
	The article is organized as follows. We begin in Sec.~\ref{Sec:Model}
	with an introduction of a continuum model of a surface of CII class
	TI. We discuss three classes of symmetry-breaking heterogeneities that
	preserve its statistical symmetry in Sec.~\ref{Sec:SBSS}, focusing on the helical network surface.  A
	single interacting helical junction is studied in Sec.~\ref{Sec:IHNM},
	and is utilized to make arguments for a localization transition in the
	helical surface network.  We conclude with experimental signatures and
	the future directions in Sec.~\ref{Sec:summary}.
	
	\section{Surface Model}\label{Sec:Model}
	
	Three dimensional TIs are characterized by symmetry-protected metallic
	surfaces that host 2D massless Dirac or Majorana quasiparticles
	\cite{Schnyder2008,Hasan2010_RMP,Qi2011_RMP}. In the absence of
	interaction, they are robust to gapping out or localization by any
	symmetry-preserving single particle scattering.  We focus on the spin
	chiral TI (class CII) \cite{Schnyder2008,Hosur2010}, characterized by
	a $\mathrm{Z}_2$ invariant. Its topologically nontrivial surface
	exhibits two-valley Dirac cones with the chemical potential pinned to
	the Dirac point. The corresponding noninteracting clean CII surface
	Hamiltonian is given by
	\begin{align}    \label{Eq:H_surf_CII_0}
	H_{0}=& v_D\int\limits_{\vex{x}}\Psi^{\dagger}
	\left[-i\hat\sigma^x\partial_x-i\hat\sigma^y\partial_y\right]\Psi,
	\end{align}
	where $\Psi$ is a four component fermionic Dirac field and
	$\hat{\sigma}^{x,y,z}$ is the ``spin'' Pauli matrix. 
	
	The clean surface Hamiltonian, $H_0$ can be perturbed by a number of
	fermion bilinear operators, $\Psi^{\dagger}\hat{T}_a\Psi$ (listed 
	in Table~\ref{Table:CII_mass}), that can be classified by
	their commutation/anticommutation with $\hat{\sigma}^x$ and $\hat{\sigma}^y$ (
	$[\hat{T}_a,\hat{\sigma}^x]$, $[\hat{T}_a,\hat{\sigma}^y]$, $\{\hat{T}_a,\hat{\sigma}^x\}$, and $\{\hat{T}_a,\hat{\sigma}^y\}$).
	If a bilinear commutes with both the $\hat{\sigma}^x$ and $\hat{\sigma}^y$, it is regarded as a scalar operator, denoted by $\hat{V}_a$. A vector operator, $\hat{A}_a$, commutes with one of the $\hat{\sigma}^x$ or $\hat{\sigma}^y$, but anticommutes with the other one. The mass operator, $\hat{M}_a$, anticommutes with both the $\hat{\sigma}^x$ and $\hat{\sigma}^y$.
	
	We first focus on the symmetric bilinear operators given by
	\begin{align}
	\label{Eq:H_surf_CII_dis}
	H_{\text{dis}}=&\int\limits_{\vex{x}}\Psi^{\dagger}
	\left[v_1\hat\tau^x+v_2\hat{\tau}^z+a_1\hat\sigma^x\hat\tau^y
	+a_2\hat\sigma^y\hat\tau^y\right]\Psi,
	\end{align}
	where $\hat{\tau}^{x,y,z}$ is the ``valley'' Pauli matrix.
	The bilinear operators $v_1(\xv)$, $v_2(\xv)$ are scalar and $a_1(\xv)$,
	$a_2(\xv)$ vector, time- and particle-hole symmetry-preserving random
	potentials\footnote{The disorder potentials here are consistent with a
		previous study, but use a different parametrization \cite{Ryu2012}.}.
	The time reversal ($\mathcal{T}$) and the particle-hole
	($\mathcal{P}$) operations are defined by
	\begin{subequations}\label{Eq:TP_CII}
		\begin{align}
		\mathcal{T}:\Psi\rightarrow &i\hat\sigma^y\Psi,\,\,\, i\rightarrow-i,\\
		\mathcal{P}:\Psi\rightarrow &\hat\sigma^x\hat\tau^y(\Psi^{\dagger})^T.
		\end{align}
	\end{subequations}
	We note that the matrices in both symmetry operations ($\hat\sigma^y$
	and $\hat\sigma^x\hat\tau^y$) are antisymmetric because they
	correspond to $\mathcal{T}^2=-1$ and $\mathcal{P}^2=-1$
	\cite{Chiu2016,Ludwig2015}. In addition, a chiral operation ($\mathcal{S}=\mathcal{T}\mathcal{P}$) can be defined as a product of $\mathcal{T}$ and $\mathcal{P}$.
	All the bilinear operators in Table~\ref{Table:CII_mass} are classified by these symmetries as well.
	
	We now consider symmetry-breaking random bilinear perturbations to the $\mathcal{T}$, $\mathcal{P}$ symmetric CII surfaces.
	Although (as listed in Table~\ref{Table:CII_mass}) there are various scalar $(\hat{V}_a)$ and vector $(\hat{A}_a)$ operators, these do not open up a gap or induce a localization, unlike the mass operator $\hat{M}_a$ \cite{Ludwig1994,Morimoto2015_AL}. We thus focus on random symmetry-breaking masses,
	$H_{\text{M}}=\sum_{a=1}^4 H_{\text{M},a}$, with
	\begin{align}\label{Eq:H_M}
	H_{\text{M},a}=\int\limits_\xv m_a(\xv)\Psi^{\dagger}\hat{M}_a\Psi.
	\end{align}
	These can be
	classified as follows (also in Table~\ref{Table:CII_mass}): $\hat{M}_1=\hat{\sigma}^z$ preserves
	$\mathcal{P}$ but breaks $\mathcal{T}$;
	$\hat{M}_2=\hat{\sigma}^z\hat{\tau}^y$ preserves $\mathcal{T}$ but
	breaks $\mathcal{P}$; $\hat{M}_3=\hat\sigma^z\hat\tau^x$ and
	$\hat{M}_4=\hat\sigma^z\hat\tau^z$ preserve $\mathcal{S}=\mathcal{T}\mathcal{P}$ but break both $\mathcal{T}$ and $\mathcal{P}$.
	
	For our purposes here, we imagine simply imposing the random
	sign-changing amplitudes, $m_a(\vex{x})$, such that statistically (averaged
	over disorder or samples) $\mathcal{T}, \mathcal{P}$ symmetries remain
	intact, i.e., $m_a$ has zero mean. More physically, such random mass operators can arise as a
	result of heterogeneous spontaneous symmetry breaking in the presence
	of symmetric quenched disorder $H_{\text{dis}}$ and four-Fermi
	interactions
	\begin{align}\label{Eq:H_I_CII}
	H_I=\sum_{a=1}^4U_a\int\limits_{\vex{x}}\left[\Psi^{\dagger}\hat{M}_a\Psi\right]^2,
	\end{align}
	where $U_a$ denotes the interaction strength corresponding to the mass
	$\hat{M}_a$ \cite{Morimoto2015_breakdown,Song2017}, with $m_a(\vex{x})$ the
	mean-field order parameter determined self-consistently
	\cite{Morimoto2015_breakdown}.

	\begin{table}[]
		
		\begin{tabular}{c|c|ccc|c}
			$\hat{T}_a$ & Billinear operator & $\mathcal{T}$ & $\mathcal{P}$ & $\mathcal{S}$ & Class\\ \hline\hline
			$\hat{V}_{1,2}$ & $\hat\tau^x$, $\hat\tau^z$ & $\checkmark$ & $\checkmark$ &$\checkmark$ & CII\\
			$\hat{A}_{1,2}$ & $\hat\sigma^x\hat\tau^y$, $\hat\sigma^y\hat\tau^y$ & $\checkmark$ & $\checkmark$ &$\checkmark$ &CII\\ \hline
			$\hat{V}_3$ & $\hat{\tau}^y$ & $\mathsf{x}$ & $\checkmark$ & $\mathsf{x}$ &C\\
			$\hat{A}_{3,4,5,6}$ & $\hat\sigma^x\hat\tau^x$, $\hat\sigma^x\hat\tau^z$, $\hat\sigma^y\hat\tau^x$, $\hat\sigma^y\hat\tau^z$ & $\mathsf{x}$ & $\checkmark$ & $\mathsf{x}$ &C\\
			$\hat{M}_1$ & $\hat\sigma^z$ & $\mathsf{x}$ & $\checkmark$ & $\mathsf{x}$ &C\\ \hline
			$\hat{V}_4$ & $\hat{1}$ & $\checkmark$ & $\mathsf{x}$ & $\mathsf{x}$ &AII\\
			$\hat{M}_2$ & $\hat\sigma^z\hat\tau^y$ & $\checkmark$ & $\mathsf{x}$ & $\mathsf{x}$ &AII\\ \hline
			$\hat{A}_{7,8}$ & $\hat\sigma^x$, $\hat\sigma^y$ & $\mathsf{x}$ & $\mathsf{x}$ &$\checkmark$ &AIII\\
			$\hat{M}_{3,4}$ & $\hat\sigma^z\hat\tau^x$,	$\hat\sigma^z\hat\tau^z$ & $\mathsf{x}$ & $\mathsf{x}$ &$\checkmark$ &AIII
		\end{tabular}
		
		\caption{Classification of the bilinear operators in terms of
			the time-reversal ($\mathcal{T}$), particle-hole
			($\mathcal{P}$), and chiral ($\mathcal{S}=\mathcal{T}\mathcal{P}$) operations on the CII class [as defined in Eq.~(\ref{Eq:TP_CII})], and the type of perturbations ($\hat{V}_a$, $\hat{A}_a$, and $\hat{M}_a$).}
		\label{Table:CII_mass}
	\end{table}
	
	Independent of the physical mechanism, we expect the CII
	symmetry-breaking random perturbation $H_{\text{M}}$ to generate a
	surface ground state that is a network of 1D domain
	walls, similar to statistical topological insulators
	\cite{Fulga2014,Morimoto2015_AL}, illustrated in
	Fig.~\ref{Fig:Domains}, the fate of which, in the presence of
	interactions is the focus of our work.

	\section{CII class symmetry-broken surface states}
	\label{Sec:SBSS}
	
	In a three-dimensional CII class TI, the random symmetry-breaking mass
	operators $\hat{M}_a$ can lead to three types of domain-wall
	networks, corresponding to three distinct symmetries of sign-changing
	masses $m_a(\xv)$ introduced in Sec.~\ref{Sec:Model} (see
	Table~\ref{Table:CII_mass}). As we will discuss below, with one type
	of a mass, the inhomogeneous symmetry breaking leads to a surface state
	composed of a network of gapless 1D domain-walls separating domains
	characterized by a positive and negative value of a mass $m_a$. In the
	CII class, it is also possible to generate multiple mass terms when  
	only the chiral symmetry ($\mathcal{S}$) is preserved. In
	this case the symmetry-breaking order parameter is a vector, that can
	rotate smoothly without vanishing, and as a result, such surface
	state, previously
	discussed \cite{Vishwanath2013,Wang2013,Wang2014,Metlitski2014,SenthilARCMP}, 
	is fully gapped.  Looking for a new, gapless but localized TI surface
	scenario, here we instead focus on the case only time-reversal ($\mathcal{T}$) or particle-hole ($\mathcal{P}$) symmetry is
	unbroken, such that there are sharp gapless domain-walls, that we will
	argue can get localized for the case of $\hat{M}_2$ disorder in the presence of interactions.
	
	The transport in such symmetry-broken surface states of CII TIs is
	governed by the resulting network of the massless 1D domain-walls.
	The domain-wall surface states can be derived analytically in the
	large domain size limit via the standard ``twist mass''
	formalism \cite{JackiwRebbi1976,SSH1979}. The 1D
	nature of the domain-walls is interrupted by regions where two domain
	walls come close to each other. These can be modeled as junctions
	illustrated in Fig.~\ref{Fig:Domains}.
	
	Here we are outlining the underlying physics and the approach, relegating the
	technical analysis to the Appendices. To make progress, we take the
	effect due to the random mass symmetry-breaking operators [given by
	Eq.~(\ref{Eq:H_M})] to be much stronger than the symmetric disorder
	[given by $H_{\text{dis}}$ in Eq.~(\ref{Eq:H_surf_CII_0})]. Therefore,
	we first compute the zero energy states of $H_0+H_{\text{M},a}$,
	determining the structure of the 1D electron domain-walls. We find
	that only one class, the helical domain-walls, arising by domains
	breaking $\mathcal{P}$ but not $\mathcal{T}$ symmetry, have the possibility of localization.
	We then study the stability of the resulting network to interactions
	and symmetric disorder, $H_{\text{dis}}$, taking advantage of our
	earlier work on 1D edges of 2D TIs \cite{Chou2018}, as well as the analysis of the
	four-way junctions \cite{Teo2009}. The other symmetry-breaking
	scenarios are robust to symmetric disorder and interactions and thus
	such disordered TI surfaces remain metallic.
	
	\subsection{Particle-hole symmetric surface: Chiral domain-wall network}
	
	A particle-hole symmetric but time-reversal broken surface corresponds
	to the random mass operator $\hat{M}_1=\hat{\sigma}^z$. In this case
	the 1D domain-walls are chiral with two co-moving electrons. The
	chiral domain-wall states can be viewed as the spin quantum Hall edge
	of class C
	\cite{Gruzberg1997,Senthil1998,Gruzberg1999,Senthil1999}. The
	intersections or proximity of chiral domain-walls can only rearrange
	their connectivity, but cannot stop the network state from conducting.
	Such a metallic state can be realized as a statistical topological
	insulator \cite{Fulga2014,Morimoto2015_AL}, or, alternatively can be
	viewed as a critical state at the plateau transition
	\cite{Evers2008}. These are well known to be robust against local
	symmetric disorder perturbations, as with conventional quantum Hall
	states. We are not aware of any new physics to be discovered here from
	the interplay of disorder and interactions, at least in the large
	domain size limit, where the domain-wall structure can be derived
	analytically.
	
	\subsection{Time-reversal symmetric surface: Helical domain-wall network}
	
	\begin{figure}[t!]
		\includegraphics[width=0.45\textwidth]{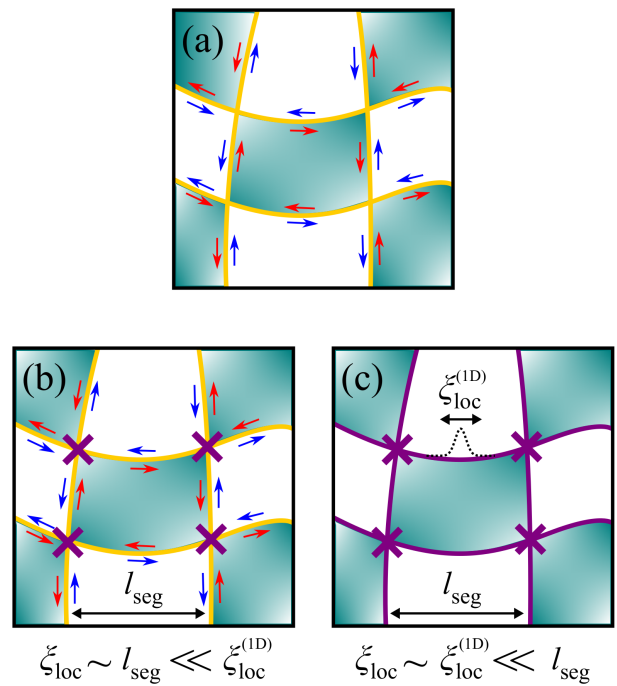}
		\caption{Three possible regimes in the helical domain-wall
			networks. (a) Metallic regime. The helical network remains
			conducting for weaker Luttinger liquid interaction, $K > 1/2$. (b) ``Clogged'' regime. For $3/8 < K
			\le 1/2$, the symmetric disorder remains irrelevant, but the
			random junctions (intersections) of the helical domain-walls
			are relevant and therefore block the dc transport, with
			helical electrons confined in the inter-junction domain-wall segments, breaking
			time-reversal symmetry spontaneously \cite{Teo2009}. A clogged state also persists for $K<3/8$ when the 1D domain-wall localization length ($\xi_{\text{loc}}^{(\text{1D})}$) is much longer than the typical length of the domain-wall segment ($l_{\text{seg}}$). As indicated in the figure, in the clogged regime, the true physical localization length, $\xi_{\text{loc}}$ is set by $l_{\text{seg}}$. 
			(c)
			``Fully-localized'' regime. For $K\le 3/8$ and sufficiently small 1D domain-wall
			localization length ($\xi_{\text{loc}}^{(\text{1D})}\ll l_{\text{seg}}$), the whole domain-wall network becomes
			localized with a localization length set by $\xi_{\text{loc}}^{(\text{1D})}$. The yellow (purple) solid lines indicate the
			conducting (localized) channels; the purple crosses mark the
			perfect barrier junctions; the blue and red arrows indicate
			the movement of the domain-wall electrons that form Kramers
			pairs in each domain-wall segment.  }
		\label{Fig:helical}
	\end{figure}
	
	We now turn to the most interesting case with a time-reversal
	symmetric surface, but with particle-hole symmetry randomly broken by
	the mass operator $\hat{M}_2=\hat{\sigma}^z\hat{\tau}^y$. In this
	case, the domain-walls form a helical network state \cite{Potter2017},
	protected against localization in the absence of interactions
	\cite{Kane2005_1,Xie2016}, and have been studied
	previously \cite{Obuse2007,Obuse2008,Ryu2010,Obuse2014}. We emphasize
	that class CII TI is the only ten-fold way insulator that realizes a
	network of helical states under inhomogeneous symmetry breaking.  The
	surface remains metallic as long as the time-reversal symmetry is
	intact.  We next discuss the stability of this metallic helical
	network to interactions and symmetry-preserving disorder in the
	remainder of this subsection, with the technical analysis presented in
	Sec.~\ref{Sec:IHNM}.
	
	Such surface transport is governed by the network of interacting
	helical domain-walls. At length scales shorter than the distance
	between junctions the physics is controlled by isolated helical domain
	walls, analyzable as a helical Luttinger liquid \cite{Wu2006,Xu2006}. For sufficiently
	strong repulsive interactions, $K < 3/8$, these can be localized
	\cite{Wu2006,Xu2006,Chou2018} due to an interplay of symmetric
	disorder and umklapp four-fermion interaction \cite{Chou2018}. Such a
	localized state spontaneously and inhomogeneously breaks the
	time-reversal symmetry and is best viewed by a localized insulator of
	$e/2$-charge Luther-Emery fermions \cite{Chou2018}. Thus for $K < 3/8$,
	such TI surface becomes a network of localized one-dimensional
	insulators. This picture is self-consistent as long as the
	localization length along the one-dimensional domain-walls is short
	compared to the typical distance between junctions of the network, a
	condition that can be satisfied by taking the domains to be
	sufficiently large.
	
	In the complementary regime of weaker interactions, $K > 3/8$, the
	isolated 1D domain-wall segments are {\it not} localized, requiring an
	analysis of the full network, controlled by domain-walls proximity
	(intersections), that we model as four-way junctions. The latter
	problem is related to the earlier studies of the corner junction
	\cite{Hou2009} and the point contact \cite{Teo2009}.  We perform a
	complementary analysis based on two helical Luttinger liquids with
	symmetry-allowed impurity perturbations in Sec.~\ref{Sec:IHNM} and
	Appendix~\ref{App:Helical_junction}. As we will demonstrate, for sufficiently strong
	interactions, $K < 1/2$, the junctions become strong impurity barriers
	that suppress all conduction (before, i.e., for weaker interaction
	than localization of isolated domain-walls, $K < 3/8$
	\cite{Wu2006,Xu2006,Chou2018}), and break the time-reversal symmetry
	spontaneously. Our results are consistent with the earlier finding in
	the helical liquid point contact study with spin-orbit couplings
	\cite{Teo2009}.
	
	Combining the results from both the junction and the domain-wall
	states, we conclude the existence of three regimes (summarized in
	Fig.~\ref{Fig:helical}) in the large domains limit. For weak
	interactions $(K>1/2)$, the helical network remains conducting and can
	be viewed as a statistical TI surface
	\cite{Fulga2014,Morimoto2015_AL}. For intermediate interactions $(3/8
	< K \le 1/2)$, the junctions break time-reversal symmetry
	spontaneously and suppress the conduction. The domain-wall state in
	each segment remains ``delocalized'', but the junctions block
	transport. We refer to this as a ``clogged'' regime.  For sufficiently
	strong interactions $(K\le 3/8)$, all the junctions and the
	domain-wall segments break time-reversal symmetry spontaneously and
	form a network of localized one-dimensional channels. Because the
	``clogged'' and ``fully localized'' states are qualitatively the same,
	they are two distinct regimes connected by a smooth crossover (driven
	by interaction strength $K$) within a single insulating phase that
	sets in for $K < 1/2$.  We discuss this crossover further in
	Sec.~\ref{Sec:IHNM}.
	
	\subsection{Surface with only chiral symmetry: Gapped insulator}
	
	Lastly, for completeness, we discuss the CII TI surface with both
	time-reversal and particle-hole symmetry broken by two mass operators,
	$\hat{M}_3=\hat{\sigma}^z\hat{\tau}^x$ and
	$\hat{M}_4=\hat{\sigma}^z\hat{\tau}^z$, corresponding to the chiral symmetric
	class AIII. 
	Qualitatively distinct from the case of a single mass, such symmetry
	broken surface state is typically fully gapped because the domains
	with multiple masses can deform from one to another without closing
	the gap \cite{Morimoto2015_AL}, a possibility that was anticipated in
	the previous studies
	\cite{Vishwanath2013,Wang2013,Wang2014,Metlitski2014,SenthilARCMP}.
	Thus, such a surface is a fully gapped insulator up to
	disorder-induced rare in-gap states. 
	
	Finally we note that for a fine-tuned microscopic model, where only
	one type of bilinear appears, e.g.,
	$\hat{M}_3=\hat{\sigma}^z\hat{\tau}^x$ or
	$\hat{M}_4=\hat{\sigma}^z\hat{\tau}^z$, a domain-wall network can be
	realized. However, the domain-walls of this network carry
	conventional one dimensional electrons. They thus do not enjoy the
	protection of $\mathcal{T}$ symmetry against localization and can therefore be
	Anderson-localized by disorder alone, in the absence of interactions.
	
	\section{Helical domain-network analysis}
	
	We now focus on a helical domain-wall network and analyze its
	stability to interactions and symmetry-preserving disorder. To this end, we first demonstrate localization along independent 1D domain-walls, and then show that the localization is stable to the ever-present domain-wall junctions, whose effect is to enhance localization by shifting the critical point to weaker interactions.
	
	\subsection{Independent helical domain-walls}
	
	At short length scales (shorter than the typical inter-junction separation) we can neglect the domain-wall junctions and focus on the nature of individual helical
	domain-wall segments. In this limit, the problem reduces to independent 1D
	helical conductors, in the presence of symmetry-preserving disorder and
	interactions. This problem is technically identical to that of
	an interacting disordered edge of a 2D TI in the AII class \cite{Wu2006,Xu2006,Chou2018}, that can be localized by the interplay of symmetric disorder and interactions.
	
	To see this, we consider a helical conductor modeled as counter-propagating states of right $(R)$ and left ($L$) moving helical fermions, with the low-energy disorder-free Hamiltonian given by
	\begin{align}\label{Eq:H:HLL1}
	H_{\text{hLL}}=v_F\int\limits_{x}\left[R^{\dagger}\left(-i\partial_xR\right)-L^{\dagger}\left(-i\partial_xL\right)\right]+H_{\text{int}},
	\end{align} 
	where $v_F$ is the Fermi velocity and $H_{\text{int}}$ encodes the Luttinger liquid interactions \cite{Giamarchi_Book,Shankar_Book}.
	Although $H_{\text{hLL}}$ takes the same form as the spinless Luttinger liquid \cite{Giamarchi_Book,Shankar_Book}, it is distinct from it, as in the helical Luttinger liquids the time-reversal symmetry ($R\rightarrow L$, $L\rightarrow - R$, and $i\rightarrow -i$) satisfies $\mathcal{T}^2=-1$, and thereby forbids single-particle backscattering perturbation, $L^{\dagger}R$. 
	
	Thus symmetric disorder only allows forward scattering, 
	\begin{align}
	H_{\text{chem}}=\int\limits_x V(x) \left(R^{\dagger}R+L^{\dagger}L\right),
	\end{align}
	that in the absence of additional interactions does not lead to localization.
	
	The helical Luttinger liquid is also generically stable to the (disorder-free) time-reversal symmetric two-particle umklapp scattering, 
	\begin{align}
	H_{\text{umklapp}}=\int\limits_{x}\left[e^{i(4k_F-Q)x}:(L^{\dagger}R)^2:+\text{H.c.}\right]
	\end{align}
	($:A:$ is the normal ordering of $A$) as long the reciprocal lattice wavevector $Q$ is sufficiently incommensurate, i.e., as long as $|4k_F-Q|>\delta Q_c$ ($\delta Q_c$ the critical threshold) is satisfied \cite{PokrovskyTalapov}.
	
	However, in the presence of symmetric disorder, that statistically makes up the wavevector incommensuration, the umklapp interaction generates a random time-reversal symmetric two-fermion back-scattering, that can lead to a localization of the 1D helical Luttinger liquid and the associated spin-glass-like time-reversal symmetry breaking \cite{Chou2018}.
	Indeed, the standard renormalization group (RG) analysis shows that an interacting
	disordered helical conductor can be localized for $K<3/8$ \cite{Wu2006,Xu2006}. Alternatively, the problem at $K=1/4$ can be mapped onto noninteracting Luther-Emery fermions \cite{Luther1974} with chemical potential disorder \cite{Chou2018}, a model that is known to give localization for the entire spectrum \cite{Bocquet1999}. Such an interacting localized state is best viewed as an Anderson localized insulator of half-charge fermions (solitons), that exhibits a nonmonotonic localization as a function of disorder strength \cite{Chou2018}.
	
	Such localization of the 1D helical liquids then directly predicts a localization of long segments of nonintersecting domain-wall, valid in the regime when domain-wall junctions can be neglected. We next analyze the complementary regime where junctions play an essential role in localization of the CII surface.
	
	\subsection{Interacting helical junction}
	\label{Sec:IHNM}
	
	For a weaker electron interaction $K > 3/8$ and/or smaller domain
	size, the domain-wall intersections become important, and it is
	necessary to take into account junctions (see zoom-in of Fig.~\ref{Fig:Domains}). At the
	technical level, the problem of the four-way helical junction is
	related to the studies of a corner junction \cite{Hou2009} and point
	contact \cite{Teo2009} in a 2D topological insulator. In
	these previous works, the junction of four semi-infinite helical
	Luttinger liquids is mapped to an infinite spinful Luttinger liquid
	with an impurity interaction. We present a technically distinct but
	physically equivalent analysis based on two isolated Luttinger liquids
	with junction perturbations.
	
	We thus consider two 1D generic helical Luttinger liquids $+, -$, interacting via a
	local junction perturbation, corresponding to two helical
	domain-walls coming to close proximity (see the zoom-in in
	Fig.~\ref{Fig:Domains}). 
	Because these are boundaries of the same type of gapped domains, they map onto two 1D Luttinger liquids of opposite helicity, described by two copies of the helical Hamiltonian [Eq.~(\ref{Eq:H:HLL1})],
	\begin{align}
	H_{\text{hLL},2}=& v_F\sum_{s=\pm}\int\limits_{x}\left[R_s^{\dagger}\left(-i\partial_x R_s\right)-L_s^{\dagger}\left(-i\partial_x L_s\right)\right]+H_{\text{int},2},
	\end{align}
	where $R_s$ ($L_s$) is the right (left) moving fermion, with the
	index $s$ labeling the two helical domain-walls and $H_{\text{int},2}$ encoding the Luttinger liquid
	interactions \cite{Giamarchi_Book} within each helical liquid. 
	For simplicity, we take these two to have the same Fermi velocity ($v_F$) and
	Luttinger liquid interaction; we expect our qualitative conclusions to remain valid away from this 
	special case. 
	%The time-reversal operation ($\mathcal{T}^2=-1$) correspond to $R_{\pm}\rightarrow L_{\pm}$, $L_{\pm}\rightarrow -R_{\pm}$, and $i\rightarrow -i$. 
	%It is important to note that there are two distinct
	%configurations setup the calculations \cite{Teo2009}. The
	%transition between the two configurations is not important for the
	%interacting junction as we will discussed in this section.
	
	To construct junction perturbations, we enumerate all bilinear and
	quartic operators allowed by the time-reversal symmetry
	\cite{Tanaka2009,Chou2015}. For example, as usual the single-particle
	backscattering within the same helical liquid ($L^{\dagger}_sR_s$) is
	forbidden.  We will also ignore perturbations that are always
	irrelevant in the RG analysis.  The single particle
	forward and backward tunneling processes between the two helical
	liquids are given by
	\begin{align}
	\nonumber H_{\text{junc}}^{(1)}=&-t_e\left[L^{\dagger}_-(0)R_+(0)-R^{\dagger}_-(0)L_+(0)+\text{H.c.}\right]\\
	\label{Eq:junction_e}&-t_{e'}\left[R^{\dagger}_-(0)R_+(0)+L^{\dagger}_-(0)L_+(0)+\text{H.c.}\right],
	\end{align}
	where $t_e$ and $t_{e'}$ are the amplitudes of single electron
	tunneling. We note that $t_{e'}$ process is only allowed in the
	presence of Rashba spin-orbit coupling, which breaks the nongeneric
	spin $S_z$ conservation \cite{Teo2009}. For sufficiently strong $t_e$,
	the connectivity of the two helical liquids may be restructured. (See
	the zoom-in of Fig.~\ref{Fig:Domains} for the two possible
	configurations.)
	
	We also include the two-particle ``Cooper pair'' tunneling processes, given
	by
	\begin{align}\label{Eq:junction_2e}
	H_{\text{junc}}^{(2)}=-t_{2e}\left[L^{\dagger}_-(0)R^{\dagger}_-(0)R_+(0)L_+(0)+\text{H.c.}\right],
	\end{align}
	corresponding to a Kramers pair hopping between two helical domain
	walls.
	
	\begin{figure}[t!]
		\includegraphics[width=0.3\textwidth]{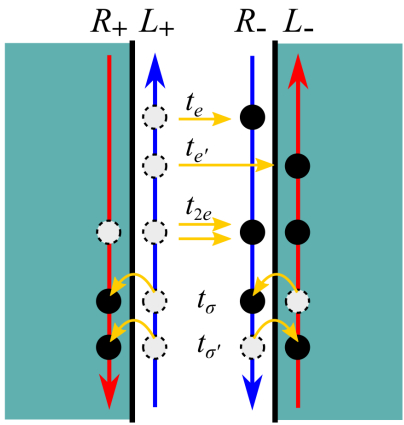}
		\caption{Illustration of junction perturbations between two
			helical domain-wall liquids with opposite helicities in
			proximity to each other. In the absence of Rashba spin-orbit coupling, the red (blue) arrows indicate movers with up (down) spin. The corresponding fermion fields are labeled on the top of this illustration. The interactions $t_e$, $t_{2e}$, and $t_{\sigma}$ are ``spin-preserving'' processes;
			$t_{e'}$ and $t_{\sigma'}$
			perturbations correspond to ``spin-flip'' processes which
			require Rahsba spin-orbit coupling.  }
		\label{Fig:processes}
	\end{figure}
	
	Finally, we include the two-particle backscattering across the
	junction,
	\begin{align}
	\nonumber H_{\text{junc}}^{(I)}=&-t_{\sigma}\left[L^{\dagger}_+(0)R_+(0)L^{\dagger}_-(0)R_-(0)+\text{H.c.}\right]\\
	\label{Eq:junction_H_I}&-t_{\sigma'}\left[L^{\dagger}_+(0)R_+(0)R^{\dagger}_-(0)L_-(0)+\text{H.c.}\right].
	\end{align} 
	The $t_{\sigma}$ and $t_{\sigma'}$ processes can be viewed as
	``spin-flip'' processes. In particular, $t_{\sigma'}$ operator breaks
	the nongeneric spin $S_z$ conservation \cite{Teo2009}. These two
	interactions are analogous to the primary inter-edge interactions in
	the studies of helical liquid drag
	\cite{Chou2015,Chou2018_TIDrag}. When $t_{\sigma}$ and $t_{\sigma'}$
	are both relevant, the junction becomes a barrier that suppresses
	electrical conduction and breaks time-reversal symmetry
	\cite{Teo2009}.
	
	In the presence of time-reversal symmetry one can also consider an
	interaction-assisted backscattering \cite{Chou2015}
	\begin{align}
	\nonumber H_{\text{junc}}^{(\text{irr})}=&-W'_+\Big[R^{\dagger}_-(0)L_-(0)R^{\dagger}_+(0)R_+(0)\\
	\nonumber&\hspace{8mm}-L^{\dagger}_-(0)R_-(0)L^{\dagger}_+(0)L_+(0)+\text{H.c.}\Big]\\
	\nonumber&-W'_-\Big[R^{\dagger}_+(0)L_+(0)R^{\dagger}_-(0)R_-(0)\\
	&\hspace{8mm}-L^{\dagger}_+(0)R_+(0)L^{\dagger}_-(0)L_+(0)-\text{H.c.}\Big].
	\end{align}
	However, standard RG analysis shows that it and all other
	perturbations are irrelevant. Thus, in the remaining discussion we
	will focus on
	$H_{\text{hLL},2}+H_{\text{junc}}^{(1)}+H_{\text{junc}}^{(2)}+H_{\text{juct}}^{(I)}$,
	processes, summarized in Fig.~\ref{Fig:processes}.
	
	To study the above processes in the presence of Luttinger liquid
	interactions, we employ standard bosonization
	\cite{Shankar_Book,Giamarchi_Book} of the above Hamiltonian. With the
	detailed derivation relegated to Appendix~\ref{App:Helical_junction},
	below we summarize the results of the leading order renormalization
	group analysis, with the RG flow equations given by
	\begin{subequations}\label{Eq:junction_RG}
		\begin{align}
		\frac{dt_e}{dl}=&\left[1-\frac{1}{2}\left(K+\frac{1}{K}\right)\right]t_e,\\
		\frac{dt_{e'}}{dl}=&\left[1-\frac{1}{2}\left(K+\frac{1}{K}\right)\right]t_{e'},\\
		\frac{dt_{2e}}{dl}=&\left[1-\frac{2}{K}\right]t_{2e},\\
		\frac{dt_{\sigma}}{dl}=&\left[1-2K\right]t_{\sigma},\\
		\frac{dt_{\sigma'}}{dl}=&\left[1-2K\right]t_{\sigma'}.
		\end{align}
	\end{subequations}
	These are consistent with the previous works on the corner junction
	\cite{Hou2009} and the quantum point contact \cite{Teo2009}. We also
	note that $t_e$ and $t_{e'}$ are at most marginal in the
	noninteracting limit, $K=1$. This ensures that the configuration of
	two helical states we consider is unchanged in the repulsive
	interaction regime. The Cooper pair tunneling $t_{2e}$ naturally becomes
	relevant for sufficiently strong attractive interactions ($K>2$).
	Thus, below we focus on the two-particle backscattering, $t_{\sigma}$
	and $t_{\sigma'}$, that become relevant for $K < 1/2$.
	
	For such strongly repulsive interactions, $K < 1/2$, we only need to
	consider $H_{\text{hLL},2}+H_{\text{junc}}^{(I)}$. As detailed in
	Appendix~\ref{App:Helical_junction}, the inter-domain-wall coupling
	decomposes the action into symmetric and antisymmetric sectors. In
	each sector, the action can be mapped to the Kane-Fisher model
	\cite{Kane1992PRB,Kane1992PRL} with $K\rightarrow 2K$. For
	$K<1/2$, the impurity interactions effectively cut (i.e., pin) the symmetric and antisymmetric Luttinger liquids. In
	the physical basis of two helical Luttinger liquids, the junction
	coupling creates a perfectly reflecting boundary condition which
	suppresses all conduction \cite{Teo2009}. The junction is therefore
	``clogged.'' Concomitantly, the time-reversal symmetry is broken
	spontaneously and heterogeneously by the network of junctions \cite{Teo2009}.
	
	At the critical point $K = 1/2$, the transmission across a single
	junction is nonzero and can be computed exactly by fermionizing the
	symmetric and antisymmetric sectors into a noninteracting model of
	Luther-Emery fermions. The scattering problem can then be solved
	exactly, with the physical transmission (T) and
	reflection (R) coefficients given by \cite{Teo2009},
	\begin{subequations}\label{Eq:junction_T_R}
		\begin{align}
		\text{T}=\left(\frac{2e^{|M_b|/v}}{e^{2|M_b|/v}+1}\right)^2,\\
		\text{R}=\left(\frac{e^{2|M_b|/v}-1}{e^{2|M_b|/v}+1}\right)^2,
		\end{align}
	\end{subequations}
	where $b=S,A$ indicates symmetric and antisymmetric sectors. Above,
	$M_S=t_{\sigma}/(\pi\alpha)$ and $M_A=t_{\sigma'}/(\pi\alpha)$, with
	$\alpha$ the ultraviolet length-scale cutoff. We note that the
	expressions are independent of the energy due to low-energy
	point-scattering approximation.  When $|M_b|/v\gg 1$, the transmission
	$\text{T}\approx 4 \exp(-2|M_b|/v)$.  Therefore, we conclude that the
	junction at $K=1/2$ is also clogged for $|M_b|/v\gg 1$.  The details
	of this analysis, extended beyond a point junction limit is relegated
	to Appendix~\ref{App:Helical_junction}.
	
	The above results can now be bootstrapped to characterize the helical
	network surface of clogged junctions. The surface can be viewed as a
	network of ideal helical conductors that are connected by clogged
	resistive junctions. Each clogged junction contributes incoherently a
	suppression factor $G_j \sim \exp\left(-2|M_j|/v\right)$, where $j$ is
	the junction index and $|M_j|$ is the amplitude of the effective potential. 
	The conductance is determined by the most conductive path in
	the network. We estimate the conductance by $G\sim\prod_j' G_j
	\sim\exp\left(-2\sum_j'|M_j|/v\right)$, where the summation runs over all
	the junctions in the most conductive path. Without loss of generality,
	the number of the junctions in the path is roughly $L/l_{\text{seg}}$
	($l_{\text{seg}}$ the typical length of the domain-wall
	segment). Combining the above estimates, we predict a surface conductance
	$G\sim\exp(-2\frac{\bar{M}}{v}\frac{L}{l_{\text{seg}}})$ where
	$\bar{M}$ is the averaged value of $|M_j|$. As a comparison, the
	conductance in the localized regime is $G\sim
	\exp\left(-2L/\xi_{\text{loc}}\right)$, where $\xi_{\text{loc}}$ is
	the averaged localization length. The exponentially small conductance
	of the clogged state and the absence of qualitative distinctions,
	argues that these regimes are a single localized state, separated by a
	smooth crossover, rather than a genuine phase transition.
	
	\section{Discussion and summary}\label{Sec:summary}
	
	We explored the stability of a 2D metallic surface of a 3D spin chiral
	(CII class) topological insulator to disorder and interaction. In the scenario of
	a symmetry broken surface that forms multiple statistically symmetric
	domains, we argued that the surface can realize a {\em gapless
		insulating} ground state, with two regimes - a network of 1D helical
	domain-walls interrupted by blockaded junctions (the clogged regime),
	and a network of localized 1D helical channels (the fully-localized
	regime). This gapless insulating surface state, realized only in the
	CII TI class, is a distinct scenario from the previously discussed
	possibilities of interacting TI surfaces \cite{Vishwanath2013,Wang2013,Wang2014,Metlitski2014,SenthilARCMP}.

	The gapless insulating surface of nontrivial TIs predicted here
	shares many experimental features with a 2D conventional Anderson
	insulator, exhibiting vanishing dc conductivity and nonzero
	compressibility. However, it may be distinguishable through real-space
	surface imaging (e.g., STM) by its low-energy states organized into
	the characteristic domain-wall network, quite different from the
	conventional 2D localized states. In addition, the half-charge
	excitations in the localized regime \cite{Chou2018} and the perfect
	barrier junctions \cite{Teo2009} should in principle be experimentally
	detectable via noise measurements.
	
	Finite temperature and finite frequency measurements may also be able
	to distinguish between the clogged and fully-localized regimes,
	tunable by the strength of interactions and disorder, with the latter
	controlling the domain size. In the clogged regime of the dilute
	domain-wall limit the temperature dependence of the surface transport
	is dictated by the weak junction links \cite{Kane1992PRL,Kane1992PRB}
	and should then exhibit the one-dimensional insulator dependence.  The
	ac conductivity will show a crossover frequency scale set by
	$\omega^*\sim v_F/l_{\text{seg}}$ ($l_{\text{seg}}$ the length of
	domain-wall segment) above which the ac conductivity is the same as
	that of a 1D helical liquid \cite{Kainaris2014}. In contrast, at low
	frequency ($\omega < \omega^*$) the ac conductivity should vanish due
	to the weak link barriers at the junctions. 
	
	In the fully-localized
	regime, the transport is governed by a network of one-dimensional
	localized insulators. The low temperature conductance due to a localized insulator should
	follow $G\sim e^{-2L/\xi_{\text{loc}}}$, where $\xi_{\text{loc}}$ is
	the localization length. The ac conductivity should show the Mott
	conductivity $\sigma\propto \omega^2$ \cite{Mott1968} up to
	logarithmic corrections. These two regimes are connected via a
	crossover for finite domain-wall segments and become distinct phases
	in the infinite domain-wall segment limit.
	
	In this work, we consider statistically symmetry-preserving disorder that creates inhomogeneous symmetry breaking. Such disorder may be generated due to the interplay of symmetric disorder and interaction, leading to instabilities of the dirty interacting topological surface states \cite{Foster2012,Nandkishore2013,Foster2014}. A systematic derivation of the heterogeneous spontaneous symmetry breaking in a dirty interacting TI is beyond the scope of the current work and is left to future studies.
	
	We note that the clogged state predicted here may also be realized in
	the Luttinger liquid networks of the (twisted) bilayer graphene and other
	related platforms
	\cite{Zhang2013,San-Jose2013,Hattendorf2013,Alden2013,Ju2015topological,Yin2016direct,Li2016gate,Li2017valley,Tong2017topological,Efimkin2018,Huang2018,Wu2018}. 
	If so, the clogged phenomenology predicted here may extend to those systems as well. We
	leave to future work the extension of the present analysis to six-way
	junctions, relevant in the twisted bilayer graphene systems
	\cite{Hattendorf2013,Alden2013}.

	\section*{Acknowledgments}
	
	We thank Matthew Foster, Jason Iaconis, Han Ma, Itamar Kimchi, and
	Abhinav Prem for useful discussions.  This work is supported in part
	by a Simons Investigator award from the Simons Foundation (Y.-Z.C. and
	L.R.), and in part by the Army Research Office under Grant Number
	W911NF-17-1-0482 (Y.-Z.C. and R.M.N.). Part of this work was performed
	(Y.-Z.C. and R.M.N.) at the KITP, supported by the NSF under Grant
	No. PHY-1748958.  The views and conclusions contained in this document
	are those of the authors and should not be interpreted as representing
	the official policies, either expressed or implied, of the Army
	Research Office or the U.S. Government. The U.S. Government is
	authorized to reproduce and distribute reprints for Government
	purposes notwithstanding any copyright notation herein.
	
	\appendix
	
	\section{Derivations of domain-wall states}\label{App:DW}
	
	Here we derive the low-energy domain-wall model from the
	2D surface theory encoded in
	$H_0+H_{\text{dis}}+H_{\text{M},a}$, Eqs.~(\ref{Eq:H_surf_CII_0}), (\ref{Eq:H_surf_CII_dis}), and
	(\ref{Eq:H_M}). Our strategy is to first solve $H_0+H_{\text{M},a}$ exactly,
	thereby obtaining the domain-wall states and then treat
	$H_{\text{dis}}$ as a perturbation.  The solution of
	$H_0+H_{\text{M},a}$ can be parametrized by
	$\Psi_0(x,y)=f(x)\left[\tilde\psi_1(y)\hat{v}_1+\tilde\psi_2(y)\hat{v}_2\right]$
	where $\tilde{\psi}_{1,2}(y)$ are normalized wavefunctions of $y$ and
	$\hat{v}_{1,2}$ is a four component vector.  Taking $m_a(x) =
	m_a\sgn(x)$, we find that the amplitude $f(x)$ and vectors
	$\hat{v}_{1,2}$ satisfy,
	\begin{align}\label{Eq:Twist_mass}
	-iv_D\hat{\sigma}^x\hat{v}_{1,2}\partial_xf(x)+m_a\sgn(x)\hat{M}_a\hat{v}_{1,2}f(x)=0,
	\end{align}
	which reduces to
	\begin{align}\label{Eq:Twist_mass2}
	\hat{v}_{1,2}\partial_xf(x)
	=-\frac{m_a}{v_D}\left(i\hat{\sigma^x}\hat{M}_a\right)\sgn(x)\hat{v}_{1,2}f(x).
	\end{align}
	The zero energy normalizable amplitude solution is given by
	\begin{align}
	f(x)=\sqrt{\frac{m_a}{v_D}}e^{-(m_a/v_D)|x|},
	\end{align}
	and the four component vectors satisfy
	\begin{align}
	i\hat{\sigma}^x\hat{M}_a\hat{v}_{1,2}=\hat{v}_{1,2}.
	\end{align}
	The above solution $f(x)$ describes the domain-wall profile across
	$x$, with the domain-wall chosen to run along $y$. The single
	domain-wall assumption is justified as long as its width $(v_D/m_a)$
	is much smaller than the typical domain size $w$, i.e., $w m_a/v_D \gg
	1$.
	
	To obtain the effective 1D domain-wall Hamiltonian we substitute
	$\Psi_0$ for $\Psi$ inside $H_0+H_{\text{dis}}+H_{\text{M},a}$. The
	resulting kinetic energy part of the domain-wall Hamiltonian is then
	given by
	\begin{align}
	\nonumber H_{\text{DW},0}=&\left[H_0+H_{\text{M},a}\right]_{\Psi\rightarrow \Psi_0}\\
	\label{Eq:H_DW_0}=&v_D\int dy\left[-is_1\tilde\psi_1^{\dagger}\partial_y\tilde\psi_1
	-is_2\tilde\psi_2^{\dagger}\partial_y\tilde\psi_2
	\right],
	\end{align}
	where $s_{1,2}=\hat{v}_{1,2}^{\dagger}\hat{\sigma}^y\hat{v}_{1,2}=\pm
	1$ determines the sign of velocities for the fermion fields $\psi_i$.
	%The values of $s_1$ and $s_2$ are determined by the eigenvalues of $\hat{\sigma}^y$ corresponding to eigenvectors $\hat{v}_1$ and $\hat{v}_2$. 
	The domain-wall model is chiral when
	$s_1=s_2$. We note that there is no mixing term because
	$[\hat{\sigma}^x\hat{M}_a,\hat{\sigma}^y]=0$.
	
	The disorder part of the Hamiltonian is given by
	\begin{align}
	\nonumber&H_{\text{DW,dis}}=H_{\text{dis}}\left[\Psi\rightarrow \Psi_0\right],\\
	\label{Eq:H_DW_dis}
	=&\sum_{a,b}\int dy\,\tilde\psi_a^{\dagger}\hat{v}_a^{\dagger}
	\left[\tilde{v}_1\hat\tau^x
	+\tilde{v}_2\hat{\tau}^z+\tilde{a}_1\hat\sigma^x\hat\tau^y
	+\tilde{a}_2\hat\sigma^y\hat\tau^y\right]\hat{v}_b\tilde\psi_b,
	\end{align}
	where $a,b=1,2$ are the one-dimensional fermion flavors. The 1D
	disorder bilinears, $\tilde{v}_{1}$, $\tilde{v}_{2}$, $\tilde{a}_{1}$,
	and $\tilde{a}_{2}$, correspond to their 2D disorder counter-parts,
	$v_1$, $v_2$, $a_1$, and $a_2$, respectively, related by,
	$\tilde{\mathcal{O}}(y)=\int dx f^2(x)\mathcal{O}(x,y)$ for
	$\mathcal{O}=v_1,v_2,a_1,a_2$. 
	
	We now use this set up to derive and analyze the structure of the
	chiral, helical, and (fine-tuned) non-topological domain-walls.
	
	\subsection{Chiral domain-walls}
	
	In the presence of only $\hat{M}_1=\hat{\sigma}^z$ mass operator, the time-reversal
	symmetry ($\mathcal{T}$) is broken, but the particle-hole
	($\mathcal{P}$) is preserved. The resulting symmetry-broken surface
	corresponds to the symmetry class C \cite{AltlandZirnbauer1997}.  The
	corresponding spinor equation reduces to
	$i\hat{\sigma}^x\hat{M}_1\hat{v}_{1,2}=\hat{\sigma}^y\hat{v}_{1,2}=\hat{v}_{1,2}$,
	with solutions
	\begin{align}\label{Eq:v12_M1}
	\hat{v}_1=\frac{1}{\sqrt{2}}
	\left[\begin{array}{c}
	1\\
	0\\
	i\\
	0
	\end{array}
	\right],\,\,\,
	\hat{v}_2=\frac{1}{\sqrt{2}}
	\left[\begin{array}{c}
	0\\
	1\\
	0\\
	i
	\end{array}
	\right].
	\end{align}
	
	We can then identify that
	$s_1=\hat{v}_1^{\dagger}\hat{\sigma}^y\hat{v}_1=1$ and
	$s_2=\hat{v}_2^{\dagger}\hat{\sigma}^y\hat{v}_2=1$.  Based on the
	structure in Eq.~(\ref{Eq:H_DW_0}), the domain-wall state only
	contains right-mover fermions.  Thus such a domain-wall solution
	realizes a chiral state, which corresponds to the spin quantum Hall
	edge of class C
	\cite{Gruzberg1997,Senthil1998,Gruzberg1999,Senthil1999}, and is
	robust against any local perturbation within a domain-wall.
	
	For completeness, we also construct the disorder potential on the
	domain-wall even though a chiral state is robust against such
	disorder.  Using Eqs.~(\ref{Eq:H_DW_dis}) and ~(\ref{Eq:v12_M1}), the
	effective disorder domain-wall Hamiltonian is given by
	\begin{align}
	\nonumber H_{\text{DW,dis}}^{(1)}\!=\!& \int\limits_y\!\left[\!\tilde{v}_2(y)
	(\psi^{\dagger}_1\psi_1-\psi^{\dagger}_2\psi_2)
	\!+\!\tilde{a}_2(y)(i\psi^{\dagger}_2\psi_1-i\psi^{\dagger}_1\psi_2)\!\right].
	\end{align}
	The above $\tilde{v}_2$ plays the role of an anti-symmetric chemical
	potential in the two right movers, and $\tilde{a}_y$ is an impurity
	forward scattering between two right movers, that cannot induce
	localization \cite{KaneFisher1995}.
	
	\subsection{Helical domain-walls}
	
	We now consider a symmetry-breaking mass
	$\hat{M}_2=\hat{\sigma}^z\hat{\tau}^y$. This mass bilinear breaks the particle-hole
	symmetry but preserves time-reversal symmetry. The symmetry-broken
	surface belongs to the class AII (the same as the 2D time-reversal
	symmetric $\mathrm{Z}_2$ TIs).  The corresponding spinor equation is
	$\hat{\sigma}^y\hat{\tau}^y\hat{v}_{1,2}=\hat{v}_{1,2}$ and yields
	solutions
	\begin{align}\label{Eq:v12_M2}
	\hat{v}_1=\frac{1}{2}
	\left[\begin{array}{c}
	1\\
	i\\
	i\\
	-1
	\end{array}
	\right],\,\,\,
	\hat{v}_2=\frac{1}{2}
	\left[\begin{array}{c}
	1\\
	-i\\
	-i\\
	-1
	\end{array}
	\right].
	\end{align}
	In this case, $s_1=\hat{v}_1^{\dagger}\hat{\sigma}^y\hat{v}_1=1$ and
	$s_2=\hat{v}_2^{\dagger}\hat{\sigma}^y\hat{v}_2=-1$.  According to
	Eq.~(\ref{Eq:H_DW_0}), the domain-wall movers are described by a right
	mover ($s_1=1$) and a left mover ($s_2=-1$).  In order to assess the
	effect of symmetric disorder, we construct the domain-wall disorder
	potential Hamiltonian based on Eq.~(\ref{Eq:H_DW_dis}), obtaining
	\begin{align}
	H_{\text{DW,dis}}^{(2)}=\int\limits_y\tilde{a}_2(y)\left[\tilde\psi^{\dagger}_1\tilde\psi_1+\tilde\psi^{\dagger}_2\tilde\psi_2\right].
	\end{align}
	The domain-wall disorder is controlled by a scalar potential
	$\tilde{a}_2$, corresponding to a randomly fluctuating chemical
	potential. Based on symmetry, one can also include $\hat{V}_4=\hat{1}$ in
	Table~\ref{Table:CII_mass}. This only creates correction to the
	existing random chemical potential fluctuation.  There are no
	additional bilinear operators with $\mathcal{T}^2=-1$, so we conclude
	that the domain-wall state is a helical state \cite{Potter2017} which
	is topologically protected from disorder in the absence of interactions
	\cite{Kane2005_1}.
	
	\subsection{Normal domain-wall}
	
	For certain microscopic models (e.g., fine tuning interactions such
	that only $U_3\neq 0$ or $U_4\neq 0$ appear), it is possible to
	realize only one mass term. Here, we perform the same analysis to
	derive the domain-wall states due to only
	$\hat{M}_3=\hat{\sigma}^z\hat{\tau}^x$ or
	$\hat{M}_4=\hat{\sigma}^z\hat{\tau}^z$ mass operators. In the two dimensions, the AIII class is topologically trivial. The spinor
	solutions ($\hat{v}_{1,2}$ for $\hat{M}_3$, $\hat{u}_{1,2}$ for
	$\hat{M}_4$) obey
	$\hat{\sigma}^y\hat{\tau}^x\hat{v}_{1,2}=\hat{v}_{1,2}$ and
	$\hat{\sigma}^y\hat{\tau}^z\hat{u}_{1,2}=\hat{u}_{1,2}$. The
	corresponding solutions are given by
	\begin{align}
	\hat{v}_1=\frac{1}{2}
	\left[\begin{array}{c}
	1\\
	1\\
	i\\
	i
	\end{array}
	\right],\,\,\,
	\hat{v}_2=\frac{1}{2}
	\left[\begin{array}{c}
	1\\
	-1\\
	-i\\
	i
	\end{array}
	\right],
	\end{align}
	and
	\begin{align}
	\hat{u}_1=\frac{1}{\sqrt{2}}
	\left[\begin{array}{c}
	1\\
	0\\
	i\\
	0
	\end{array}
	\right],\,\,\,
	\hat{u}_2=\frac{1}{\sqrt{2}}
	\left[\begin{array}{c}
	0\\
	1\\
	0\\
	-i
	\end{array}
	\right]
	\end{align}
	
	We thus identify that
	$\hat{v}_1^{\dagger}\hat{\sigma}^y\hat{v}_1=\hat{u}_1^{\dagger}\hat{\sigma}^y\hat{u}_1=1$
	(right mover) and
	$\hat{v}_2^{\dagger}\hat{\sigma}^y\hat{v}_2=\hat{u}_2^{\dagger}\hat{\sigma}^y\hat{u}_2=-1$
	(left mover).  Therefore, both cases give a non-chiral state. Because
	the surface state is in class A, the massless domain-wall hosts
	non-topological 1D fermions.
	
	For completeness, we also discuss the corresponding domain-wall
	disorder. With the mass $\hat{M}_3$, the disorder part is given by
	\begin{align}
	\nonumber H_{\text{DW,dis}}^{(3)}=&\int\limits_y\tilde{v}_1(y)\left[\tilde\psi^{\dagger}_1\tilde\psi_1-\tilde\psi^{\dagger}_2\tilde\psi_2\right]\\
	&+\int\limits_y\tilde{a}_1(y)\left[\tilde\psi^{\dagger}_2\tilde\psi_1+\tilde\psi^{\dagger}_1\tilde\psi_2\right].
	\end{align}
	For $\hat{M}_4$ case we instead find,
	\begin{align}
	\nonumber H_{\text{DW,dis}}^{(4)}=&\int\limits_y\tilde{v}_2(y)\left[\tilde\psi^{\dagger}_1\tilde\psi_1-\tilde\psi^{\dagger}_2\tilde\psi_2\right]\\
	&-\int\limits_y\tilde{a}_1(y)\left[\tilde\psi^{\dagger}_2\tilde\psi_1+\tilde\psi^{\dagger}_1\tilde\psi_2\right].
	\end{align}
	The antisymmetry chemical potentials ($\tilde{v}_1$ in
	$H_{\text{DW,dis}}^{(3)}$ and $\tilde{v}_2$ in
	$H_{\text{DW,dis}}^{(4)}$ ) couples to the difference of right and
	left mover local densities.  Both cases allow for conventional
	impurity backscattering ($\tilde a_1$ in both cases) within the domain
	wall, and thus realizes topologically trivial 1D fermions, which are 
	therefore not protected against Anderson localization.
	
	\section{Helical junction}\label{App:Helical_junction}
	
	In this appendix, we provide the derivations of the results in
	Sec.~\ref{Sec:IHNM}. We will also review the standard bosonization and
	the Luther-Emery analysis.
	
	\subsection{Bosonization}
	
	To treat the Luttinger interaction nonperturbatively, we adopt the standard field theoretic bosonization method \cite{Shankar_Book}. The fermionic fields can be described
	by chiral bosons via
	\begin{align}
	R_a(x)=\frac{\hat{U}_a}{\sqrt{2\pi\alpha}}e^{i\left[\phi_a+\theta_a\right](x)},\,\,L_a(x)=\frac{\hat{U}_a}{\sqrt{2\pi\alpha}}e^{i\left[\phi_a-\theta_a\right](x)},
	\end{align}
	where $\phi_{a=\pm}$ is the bosonic phase field, $\theta_{a=\pm}$ is
	the phonon-like boson, $U_{a=\pm}$ is the Klein factor
	\cite{Giamarchi_Book}, and $\alpha$ is the ultraviolet length scale
	that is determined by the microscopic model.  The time-reversal
	operation ($\mathcal{T}^2=-1$) in the bosonic language is defined as
	follows: $\phi_{\pm}\rightarrow -\phi_{\pm}+\frac{\pi}{2}$,
	$\theta_{\pm}\rightarrow\theta_{\pm}-\frac{\pi}{2}$, and
	$i\rightarrow-i$. This corresponds to the fermionic operation
	$R_{\pm}\rightarrow L_{\pm}$, $L_{\pm}\rightarrow -R_{\pm}$, and
	$i\rightarrow -i$. We note that the introduction of the Klein factors
	($U_{a=\pm}$) here is just for bookkeeping purposes.
	
	Now, we perform the standard bosonization and analyze the
	Hamiltonian. The Hamiltonian of each helical liquid is bosonized to
	\begin{align}\label{Eq:junction:H_0_b}
	H_{\text{hLL},2}=\sum_{a=\pm}\int\limits_x\left[\frac{v}{2\pi K}\left(\partial_x\theta_a\right)^2+\frac{vK}{2\pi}\left(\partial_x\phi_a\right)^2\right],
	\end{align}
	where we have assumed the same velocity ($v$) and the same Luttinger
	parameter ($K$) among the two helical liquids. $K$ encodes the
	strength Luttinger liquid interactions. $K<1$ ($K>1$) for repulsive
	(attractive) interactions. $K=1$ is at the non-interacting fermion
	limit.  The impurity perturbations [given by
	Eqs.~(\ref{Eq:junction_e}), (\ref{Eq:junction_2e}), and
	(\ref{Eq:junction_H_I})] are bosonized to
	\begin{align}
	\nonumber H_{\text{junc}}^{(1)}=&-\frac{t_e}{2\pi\alpha}\left[\hat{U}_-^{\dagger}\hat{U}_+(2i)e^{i(\phi_+-\phi_-)}\sin\left(\theta_++\theta_-\right)+\text{H.c.}\right]\\
	&-\frac{t_{e'}}{2\pi\alpha}\left[\hat{U}_-^{\dagger}\hat{U}_+
	2e^{i(\phi_+-\phi_-)}\cos\left(\theta_+-\theta_-\right)
	+\text{H.c.}\right],\\
	H_{\text{junc}}^{(2)}=&-\frac{t_{2e}}{4\pi\alpha^2}\left[\hat{U}_-^{\dagger}\hat{U}_-^{\dagger}\hat{U}_+\hat{U}_+e^{i(2\phi_+-2\phi_-)}+\text{H.c.}\right],\\
	\label{Eq:junction:H_int_b}H_{\text{junc}}^{(I)}=&-\frac{t_{\sigma}}{2\pi^2\alpha^2}\cos\left[2\theta_++2\theta_-\right]
	-\frac{t_{\sigma'}}{2\pi^2\alpha^2}\cos\left[2\theta_+-2\theta_-\right].
	\end{align}
	The corresponding renormalization group equations can be found in Eq.~(\ref{Eq:junction_RG}).
	
	\subsection{Clogged junction}
	
	We are interested in the repulsive interacting regime ($K<1$) in the
	helical network model. Therefore, we focus on the $t_{\sigma}$ and
	$t_{\sigma'}$ interactions given by Eq.~(\ref{Eq:junction:H_int_b})
	and ignore other processes. In the strong coupling limit ($K<1/2$),
	the ground state constraints are
	$\theta_+(t,x=0)+\theta_-(t,x=0)=n\pi$ and
	$\theta_+(t,x=0)+\theta_-(t,x=0)=m\pi$ where $n$ and $m$ are
	integers. The ground state yields static solutions at $x=0$:
	$\theta_+(t,x=0)=(n+m)\pi/2$ and $\theta_-(t,x=0)=(n-m)\pi/2$. As a
	consequence, the current
	$I_{\pm}=-\frac{1}{\pi}\partial_t\theta_{\pm}$ at $x=0$ is zero in
	both of the helical liquids.  Therefore, we predict that a four-way
	junction with semi-infinite helical liquids becomes ``clogged'' for
	$K<1/2$.
	
	An alternative way to view the clogging is to map the problem to a
	modified Kane-Fisher single impurity problem
	\cite{Kane1992PRL,Kane1992PRB}. We define symmetric and anti-symmetric
	collective bosonic modes as follows:
	\begin{align}
	\Theta_{S}=&\frac{1}{\sqrt{2}}(\theta_++\theta_-),\,\,\Phi_{S}=\frac{1}{\sqrt{2}}(\phi_++\phi_-),\\
	\Theta_{A}=&\frac{1}{\sqrt{2}}(\theta_+-\theta_-),\,\,\Phi_{A}=\frac{1}{\sqrt{2}}(\phi_+-\phi_-).
	\end{align} 
	The subscript $S$ and $A$ denote the symmetric and anti-symmetric
	collective modes respectively. Now, we use the collective coordinate
	to rewrite the theory. The Luttinger liquid Hamiltonian in
	Eq.~(\ref{Eq:junction:H_0_b}) is now expressed by
	\begin{align}
	H_{\text{hLL},2}=&\int\limits_x\left[\frac{v}{2\pi K}\left(\partial_x\Theta_S\right)^2+\frac{vK}{2\pi}\left(\partial_x\Phi_S\right)^2\right]\\
	&+\int\limits_x\left[\frac{v}{2\pi K}\left(\partial_x\Theta_A\right)^2+\frac{vK}{2\pi}\left(\partial_x\Phi_A\right)^2\right].
	\end{align}
	We note that the impurity interaction can not induce renormalization
	of the velocity and Luttinger parameter. The junction interactions in
	Eq.~(\ref{Eq:junction:H_int_b}) becomes to
	\begin{align}
	H_{\text{junc}}^{(I)}=-\frac{t_{\sigma}}{2\pi^2\alpha^2}\cos\left[2\sqrt{2}\Theta_S\right]
	-\frac{t_{\sigma'}}{2\pi^2\alpha^2}\cos\left[2\sqrt{2}\Theta_A\right].
	\end{align}
	Both the symmetric and anti-symmetric sectors can be individually
	mapped to the Kane-Fisher problem \cite{Kane1992PRB,Kane1992PRL} with
	$K\rightarrow 2K$. The critical point is given by $K=1/2$ below which
	the transmission of both the symmetric and anti-symmetric modes vanish
	to zero.
	
	\subsection{Luther-Emery Analysis}
	
	At the critical point $K=1/2$, one can perform standard
	refermionization for the two helical Luttinger liquids problem since
	both the symmetric and the anti-symmetric sectors correspond to the
	Kane-Fisher model \cite{Kane1992PRL,Kane1992PRB}. We introduce the
	Luther-Emery fermions via
	\begin{align}
	\nonumber\Psi_{b,R}(x)=&\frac{e^{i\left[\Phi_b(x)/\sqrt{2}+\sqrt{2}\Theta_b(x)\right]}}{\sqrt{2\pi\alpha}},\\
	\Psi_{b,L}(x)=&\frac{e^{i\left[\Phi_b(x)/\sqrt{2}-\sqrt{2}\Theta_b(x)\right]}}{\sqrt{2\pi\alpha}},
	\end{align}
	where $b=S,A$ is the index for symmetric ($S$) and antisymmetric ($A$) collective modes. The Luther-Emery fermion Hamiltonian of the sector $b$ is given by
	\begin{align}
	\nonumber H_{b}=&-iv\int dx\left[\Psi^{\dagger}_{b,R}\partial_x\Psi_{b,R}-\Psi^{\dagger}_{b,L}\partial_x\Psi_{b,L}\right]\\
	&+M_b\left[\Psi^{\dagger}_{b,R}\Psi_{b,L}+\Psi^{\dagger}_{b,L}\Psi_{b,R}\right]_{x=0},
	\end{align}
	where $M_{b=S}=t_{\sigma}/(\pi\alpha)$ and $M_{b=A}=t_{\sigma'}/(\pi\alpha)$. The impurity mass problem can be solved via standard quantum mechanical scattering approach. First, we derive the Dirac equation as follows:
	\begin{align}
	&\left[\begin{array}{cc}
	-iv\partial_x & M_b\delta(x)\\
	M_b\delta(x) & iv\partial_x
	\end{array}\right]
	\left[\begin{array}{c}
	\Psi_{b,R}\\
	\Psi_{b,L}
	\end{array}
	\right]=E
	\left[\begin{array}{c}
	\Psi_{b,R}\\
	\Psi_{b,L}
	\end{array}
	\right]\\
	\rightarrow& -iv\hat{\sigma}^z\partial_x\hat\Psi_b+M_b\delta(x)\hat{\sigma}^x\hat{\Psi}_b=E\hat{\Psi}_b,
	\end{align}
	where $\hat{\Psi}_b$ is the two-component column vector that contains $\Psi_{b,R}$ and $\Psi_{b,L}$.
	The above equation satisfies a boundary condition as follows:
	\begin{align}
	-iv\hat{\sigma}^z\left[\hat{\Psi}_b(0^+)-\hat{\Psi}_b(0^-)\right]+M_b\hat{\sigma}^x\hat{\Psi}_b(0)=0.
	\end{align}
	We note that this boundary condition is ambiguous because the wavefunction might be discontinuous at $x=0$.\\
	
	\begin{figure}[t!]
		\includegraphics[width=0.475\textwidth]{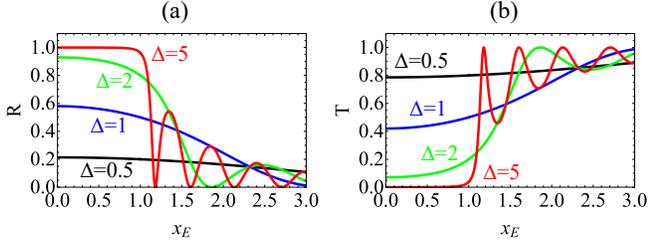}
		\caption{Reflection and transmission of 1D Dirac scattering
			problem (finite mass region) as functions of rescaled
			energy. (a) Reflection, $\text{R}=|A|^2$. (b) Transmission,
			$\text{T}=|D|^2$. Both $A$ and $B$ are given by
			Eq.~(\ref{Eq:junction_ABCD}). $x_E$ is the dimensionless
			energy parameter defined in the text below
			Eq.~(\ref{Eq:junction_ABCD}).  Black, blue, green, and red
			curves indicate $\Delta=M_b/v=0.5,1,2,5$ respectively.  The
			perfect transmissions (T$=1$ and R$=0$) for $x_E>1$
			correspond to the Fabry-P\'erot interference.  }
		\label{Fig:RT}
	\end{figure}
	
	Instead of studying the delta distribution problem, we replace
	the impurity potential by a square well potential, $M_b\delta(x)\rightarrow
	\tilde{M}_b\Theta(x)\Theta(d-x)$, where $d$ is the size of mass region
	and $\tilde{M}_b=M_b/d$ is the ``mass'' strength. The impurity
	limit is obtained by taking $d\rightarrow 0^+$.  With a finite $d$,
	the wavefunction is continuous everywhere because of the
	analyticity. We consider a scattering ansatz as follows:
	\begin{align}
	\hat{\Psi}_b(x)\!=\!\begin{cases}
	e^{ikx}\left[\begin{array}{c}
	1\\
	0
	\end{array}
	\right]+Ae^{-ikx}\left[\begin{array}{c}
	0\\
	1
	\end{array}
	\right],&\text{for }x\le 0,\\[4mm]
	Be^{iqx}\!\!
	\left[\!\!\begin{array}{c}
	1\\
	\frac{-vq+E}{\tilde{M}_b}
	\end{array}\!\!
	\right]\!
	+\!Ce^{-iqx}\!\!\left[\!\!\begin{array}{c}
	\frac{-vq+E}{\tilde{M}_b}\\
	1
	\end{array}\!\!
	\right],&\text{for }0\!<\!x\!\le\! d,\\[4mm]
	De^{ikx}\left[\begin{array}{c}
	1\\
	0
	\end{array}
	\right],&\text{for }x>d,
	\end{cases}
	\end{align}
	where $k=E/v$ and $q=\sqrt{E^2-\tilde{M}_b^2}/v$. The boundary conditions are given by
	\begin{align}
	B+C\left(\frac{-vq+E}{\tilde{M}_b}\right)=&1,\\
	B\left(\frac{-vq+E}{\tilde{M}_b}\right)+C=&A,\\
	Be^{iqd}+Ce^{-iqd}\left(\frac{-vq+E}{\tilde{M}_b}\right)=&De^{ikd},\\
	Be^{iqd}\left(\frac{-vq+E}{\tilde{M}_b}\right)+Ce^{-iqd}=&0.
	\end{align}
	With the help of Mathematica, one can obtain the solutions as follows:
	\begin{subequations}\label{Eq:junction_ABCD}
		\begin{align}
		A=&
		\frac{\left(\sqrt{x_E^2-1}-x_E\right) \left(-1+e^{2i\Delta\sqrt{x_E^2-1}}\right)}{1+\left(2 x_E\sqrt{x_E^2-1}-2 x_E^2+1\right) e^{2i\Delta\sqrt{x_E^2-1}}},\\
		B=&\frac{1}{1+\left(2 x_E \sqrt{x_E^2-1}-2 x_E^2+1\right) e^{2i\Delta\sqrt{x_E^2-1}}},\\
		C=&\frac{ \left(\sqrt{x_E^2-1}-x_E\right) e^{2i\Delta\sqrt{x_E^2-1}}}{1+\left(2 x_E \sqrt{x_E^2-1}-2 x_E^2+1\right) e^{2i\Delta\sqrt{x_E^2-1}}},\\
		D
		=&\frac{2 \left(x_E \sqrt{x_E^2-1}-x_E^2+1\right) 
			e^{i\Delta\left(\sqrt{x_E^2-1}-x_E\right)}}{1+\left(2 x_E \sqrt{x_E^2-1}-2 x_E^2+1\right) 
			e^{2i\Delta\sqrt{x_E^2-1}}},
		\end{align}
	\end{subequations}
	where $x_E\equiv E/|\tilde{M}_b|$ and $\Delta\equiv
	d|\tilde{M}_b|/v=|M_b|/v$. The reflection is $\text{R}=|A|^2$ and
	transmission is $\text{T}=|D|^2$. The dependence of $x_E$ and $\Delta$
	are plotted in Fig.~\ref{Fig:RT}. For $\Delta=|M_b|/v\gg 1$, the
	scattering problem reveals a sharp gap structure because
	$\text{R}\approx 1$ for $x_E<1$. For $x_E>1$, there are some special
	energies that allow perfect transmission. This is related to the
	Fabry-P\'erot interference. However, we do not focus on such high
	energy phenomenon in this work.
	
	Now, we consider $d\rightarrow 0^+$ with $\tilde{M}_b d=M_b$
	fixed. The finite mass region is reduced to a single impurity potential. In
	the impurity case, $x_E=Ed/|M_b|\rightarrow 0$ for a fixed $M_b/v$. The
	expression of transmission and reflection are reduced to
	Eq.~(\ref{Eq:junction_T_R}).  The results do not depend on the energy
	due to the infinite $|\tilde{M}_b|=|M_b|/d$ in this limit. These
	results characterize the low energy scattering in the network model.
	In particular, the transmission $\text{T}\rightarrow 4e^{-2|M_b|/v}$
	when $|M_b|/v\gg 1$.
	
	In the four-way junction problem, the clogging conditions at $K=1/2$
	correspond to perfect reflections in both the symmetric and
	antisymmetric sectors. In the zero energy limit, the clogging
	conditions are $|M_S|/v\gg 1$ and $|M_A|/v\gg 1$.  To make the
	junction more realistic, we can assume that both the domain-wall
	segment and the interacting region are finite. The longest wavelength
	is set by the typical domain-wall segment length, $l_{\text{seg}}$,
	corresponding to the lowest kinetic energy
	$E_0=v({2\pi}/{l_{\text{seg}}})$. The clogging conditions become to
	$v({2\pi}/{l_{\text{seg}}})<|t_{\sigma}|/(\pi \alpha d)$,
	$v({2\pi}/{l_{\text{seg}}})<|t_{\sigma'}|/(\pi \alpha d)$,
	$|t_{\sigma}|/(v\pi \alpha )\gg 1$, and $|t_{\sigma'}|/(v\pi \alpha
	)\gg 1$. The former two conditions are from comparing the energy of
	the electron to the local mass; the latter two conditions are related
	to the existence of sharp gaps.
	
	%%\bibliography{TI_ref}
	
	%%%%%%%%%%%%%
	

	%%%%%%%%%%%%%%%%
	
\end{document}